\documentstyle[11pt,ysc,twoside,epsf,graphicx]{article}
\def\edcomment#1{\iffalse\marginpar{\raggedright\sl#1\/}\else\relax\fi}
\reversemarginpar

\begin{document}
\title{Diagnostics of SS433 with the RXTE}

\author{Filippova E., Revnivtsev M.}
\affil{Space Research Institute, Russian Academy of Sciences,
              Profsoyuznaya 84/32, 117997 Moscow, Russia}

\author{Fabrika S.}
\affil{ Special Astrophysical Observatory, Nizhnij Arkhyz,
Karachaevo-Cherkesiya, 369167,  Russia}

\author{Postnov K., Seifina E.}
\affil{Sternberg Astronomical Institute, 119992, Moscow, Russia}

\begin{abstract}
We present analysis of extensive monitoring of SS433 by the RXTE
observatory collected over the period 1996-2005. The difference
between energy spectra taken at different precessional and orbital
phases shows the presence of a strong photoabsorption near the
optical star, probably due to its powerful dense wind. Assuming that
a precessing accretion disk is thick, we recover the temperature
profile in the X-ray emitting jet that best fits the observed
precessional variations of the X-ray emission temperature. The
hottest visible part of the X-ray jet is located at a distance of
$l_0/a\sim0.06-0.09$, or $\sim2-3\times10^{11}$cm from the central
compact object and has a temperature of about $T_{\rm max}\sim30$
keV. We discovered appreciable orbital X-ray eclipses at the
``crossover'' precessional phases (jets are in the plane of the sky,
disk is edge-on) which put a lower limit on the size of the optical
component  $R/a\ga0.5$ and an upper limit on a mass ratio of binary
companions $q=M_{\rm x}/M_{\rm opt}\la0.3-0.35$. The size of the
eclipsing region can be larger than secondary's Roche lobe because
of substantial photoabsorption by dense stellar wind. This must be
taken into account when evaluating the mass ratio from analysis of
X-ray eclipses.
\end{abstract}

\section{Introduction}

SS 433 is a binary system which is thought to consist of a compact
object (probably a black hole) and optical star filling its Roche
lobe.

There is a supercritical accretion rate in the system (Fabrika
2004), which leads to formation of the geometrically thick accretion
disk around compact object and two oppositely directed highly
collimated jets.

The main contribution to the observed X-ray emission from the system comes from the
optically thin multitemperature plasma in the jets via bremsstrahlung mechanism.

The X-ray emission from SS433 is subjected to systematic variations
due to the thick accretion disk precession ($P_{\rm prec}=162.375$
days), nutation ($P_{\rm nut}=6.2877$ days)
 and orbital ($P_{\rm orb}=13.08211$ days) motion in
the binary system (Fabrika 2004).

\section{The inner jet tomography by orbital and precessional eclipses}

Eclipses by the thick accretion disk and optical star result in
obscuring the innermost the hottest parts of the jets from the
observer which leads to the drop of the maximum temperature of the
registered X-ray emission (Fig.1).

\begin{figure}[htb]
\begin{center}
\includegraphics[width=0.6\columnwidth,bb=33 186 570 654,clip]{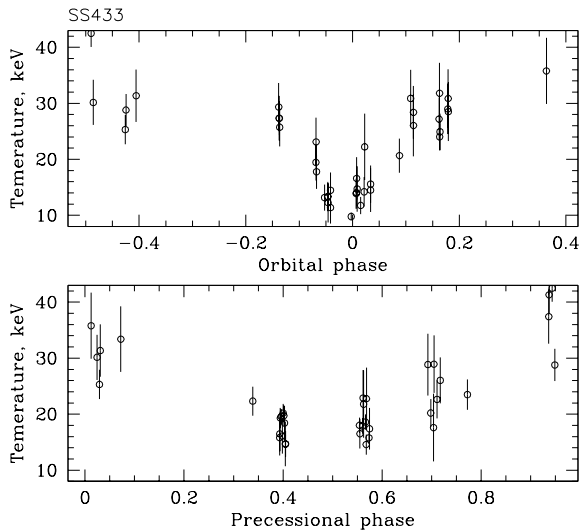}
\end{center}
\caption{The temperature of optically thin thermal plasma
emission observed from SS433 as a function of orbital and precessional phases.
 For the plot in the bottom panel only
off-eclipse data were selected.}
\end{figure}

In order to find the spectral contribution from the innermost (the
hottest) regions of the jet we examine differences
between spectra taken during eclipses and off eclipses.

The result is presented in the Fig.2. A strong photoabsorption is
observed near the orbital eclipse (the left panel of Fig.2). The
best-fit value of the absorption column density in this spectrum is
$N_{\rm H}=(12.5\pm1.5)\times 10^{22}$ cm$^{-2}$, much higher than
the conventional value.

The spectrum of the hottest (innermost) part of the jet as derived
from precessional variations off the primary eclipse (the right
panel of Fig.2) also suggests some photoabsorption, but
of considerably smaller value -- here the best-fit absorbing column density
is $N_{\rm H}=(4.5\pm1.5)\times 10^{22}$ cm$^{-2}$.

So our investigation shows the presence of the absorbing material in
the system.  Its density is much higher near the optical star. The
ordinary of this material can appear from powerful winds from the
optical supergiant star and super-Eddington accretion disk.
\begin{figure}
\plottwo{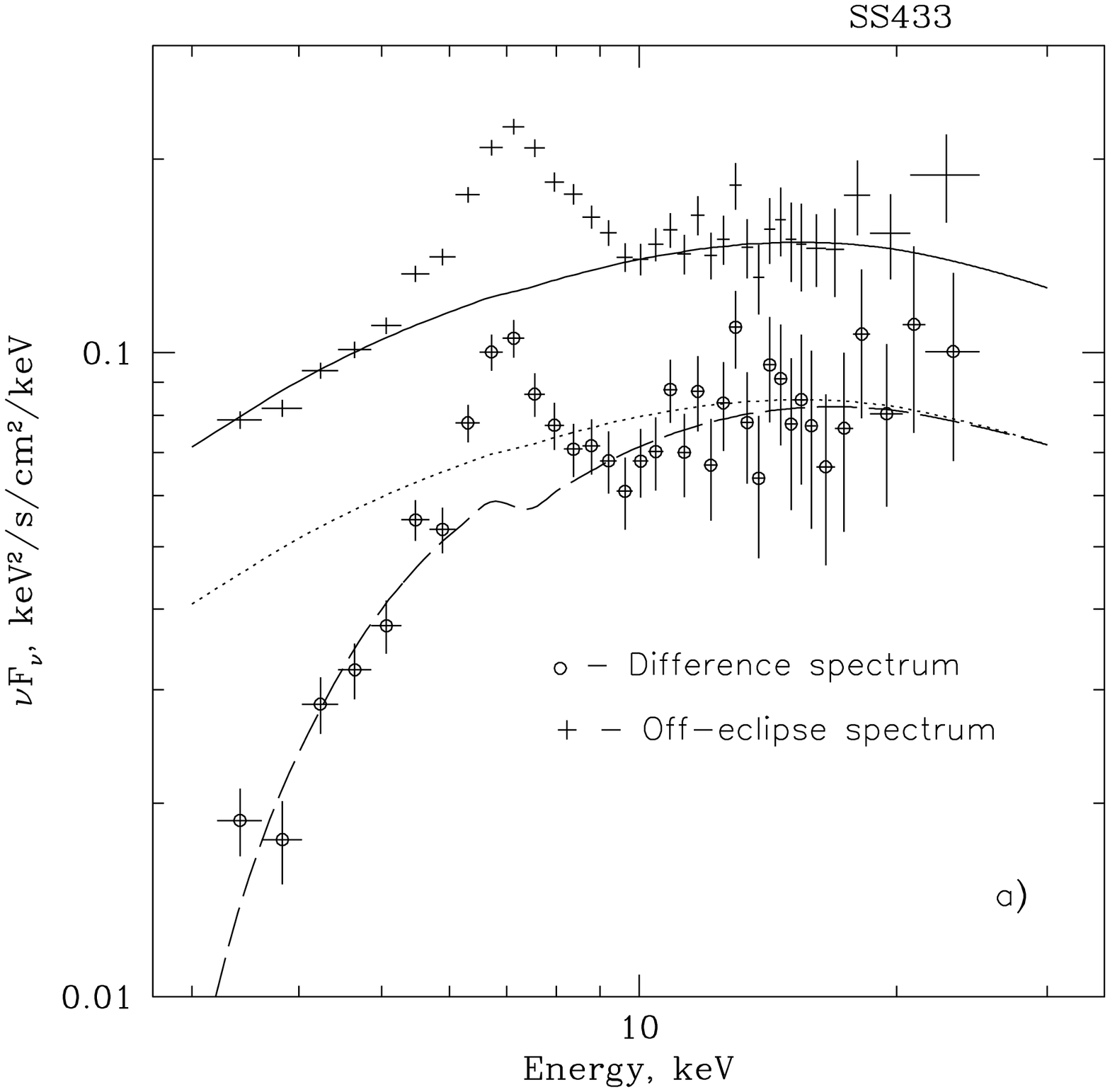}{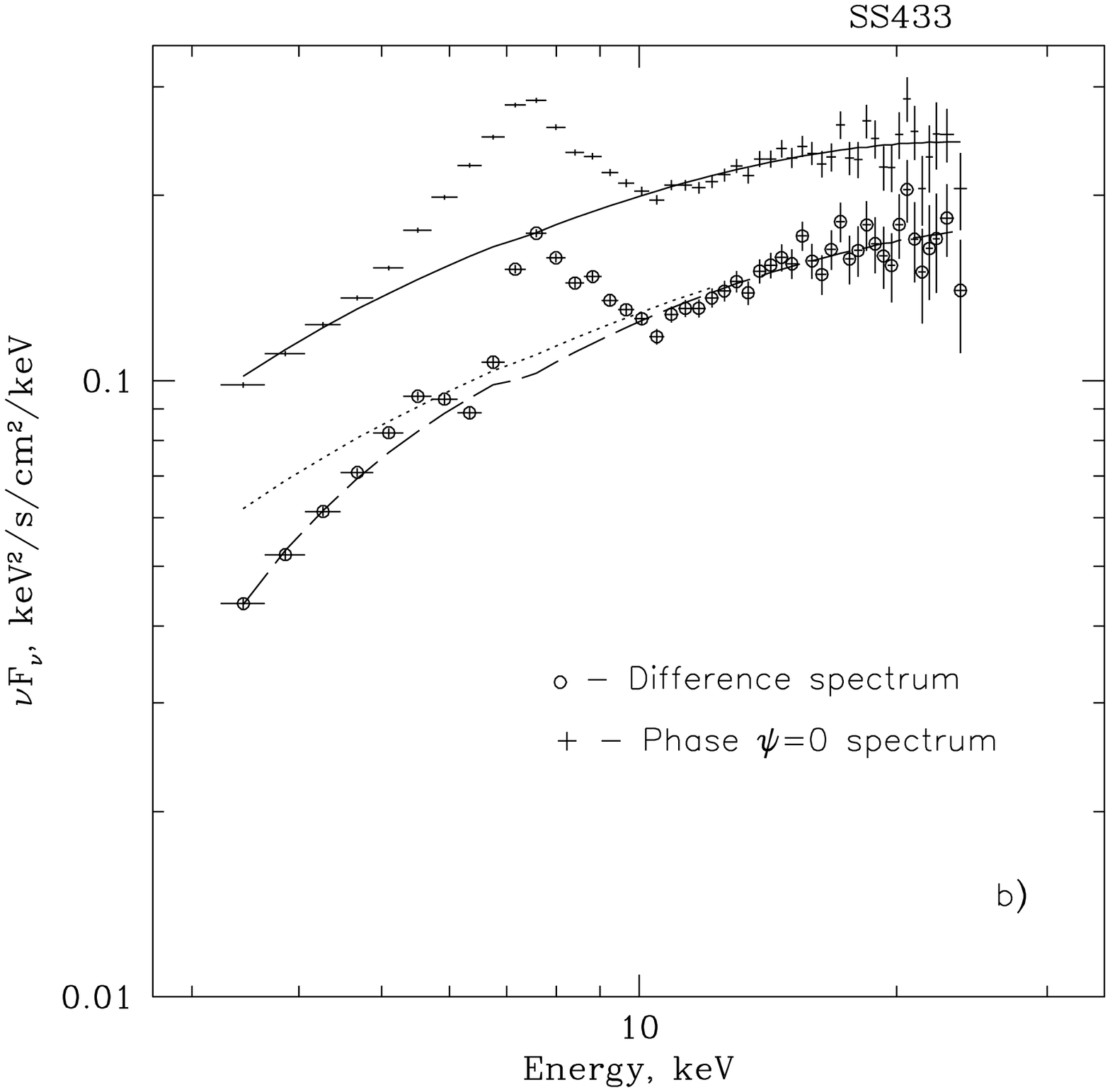}
\caption{ a) The X-ray spectrum of SS433 immediately after the eclipse
(the orbital phase $\phi=0.114$, crosses) and difference between this spectrum
and that at $\phi=0.021$. b) The X-ray spectrum of SS433 at
precessional phase $\psi\sim0$ and difference between this spectrum
and that at $\psi\sim0.4$. Solid curves show
best-fit thermal bremsstrahlung model with photoabsorption. Dotted curves show best-fit
bremsstrahlung models  with a nominal
photoabsorption column density of $N_{\rm H}=10^{22}$ cm$^{-2}$.}
\end{figure}

In order to estimate the possible increase of the star radius due to the
dense stellar wind
we take parameters of the wind (its mass loss rate and velocity distribution)
from the work of Achmad et al. (1997).
In the case of  the plasma with solar abundance it becomes almost opaque for X-rays
 when $ N_H=10^{24} cm^{-2}$.
Our estimations show that maximal value of the A supergiant mass
loss rate $\dot{M}=8\times10^{-7}-10^{-6}M_{\sun}/yr$ can increase
the size of the star up to 20 \%.

\section{Jet eclipses by the thick disk}

We describe jets in SS433 as conical plasma flows with constant velocity
$v_j=0.26\,c$ along the jet axis and constant half-opening angle $\theta=0.61^{\circ}$
(Marshall et al. 2002).
 Radius of the jet cross section
is $r=r_{b}+\theta l$, where $r_b$ is the radius of the jet near the
compact object \footnote{This analytical description does not mean
that the jet should start near the compact object.}, $l$ is the
distance from the compact object along jet axis.

\begin{figure}[htb]
\begin{center}
\includegraphics[width=0.6\columnwidth,bb=48 185 565 519,clip]{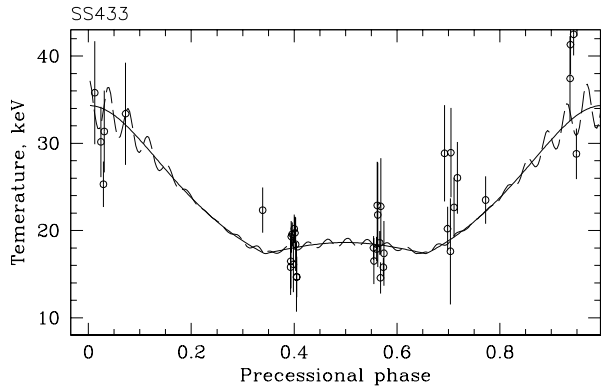}
\caption{The plasma temperature as a function of the precessional phase. The solid curve
shows the best-fit model. The dashed curve shows the same model but
with nodding motion of thick accretion disk included.}
\end{center}
\end{figure}

We assume that the dominant cooling process of the jet is adiabatic
cooling.

So the plasma temperature $T$ changes with distance from the
compact object $l$ as: $T/T_0=(1+\theta (l-l_0)/r_0)^{-4/3}$,
where $T_0$ is the plasma temperature
and $r_0$ is the jet cross section radius
at some distance $l_0$ from the central source (see also Koval' \& Shakura 1989).

For disk radius we adopted the disk truncation
radius given by Paczynski (1977).

Comparison of
the highest visible temperature profiles derived above with
observational data enables us to find the model parameters (the best-fit model is shown in Fig.3).
The best-fit parameters of our model referred to the distance
from the black hole $l_0/a=0.06-0.09$ (depending on the
assumed value of $q$) are:  the highest
temperature of the jet $T_0=30\pm2$ keV, the radius of the jet
$r_0/a=(1-1.6)\times10^{-2}$.
\section{Eclipses by the optical companion}

\begin{figure}
\plottwo{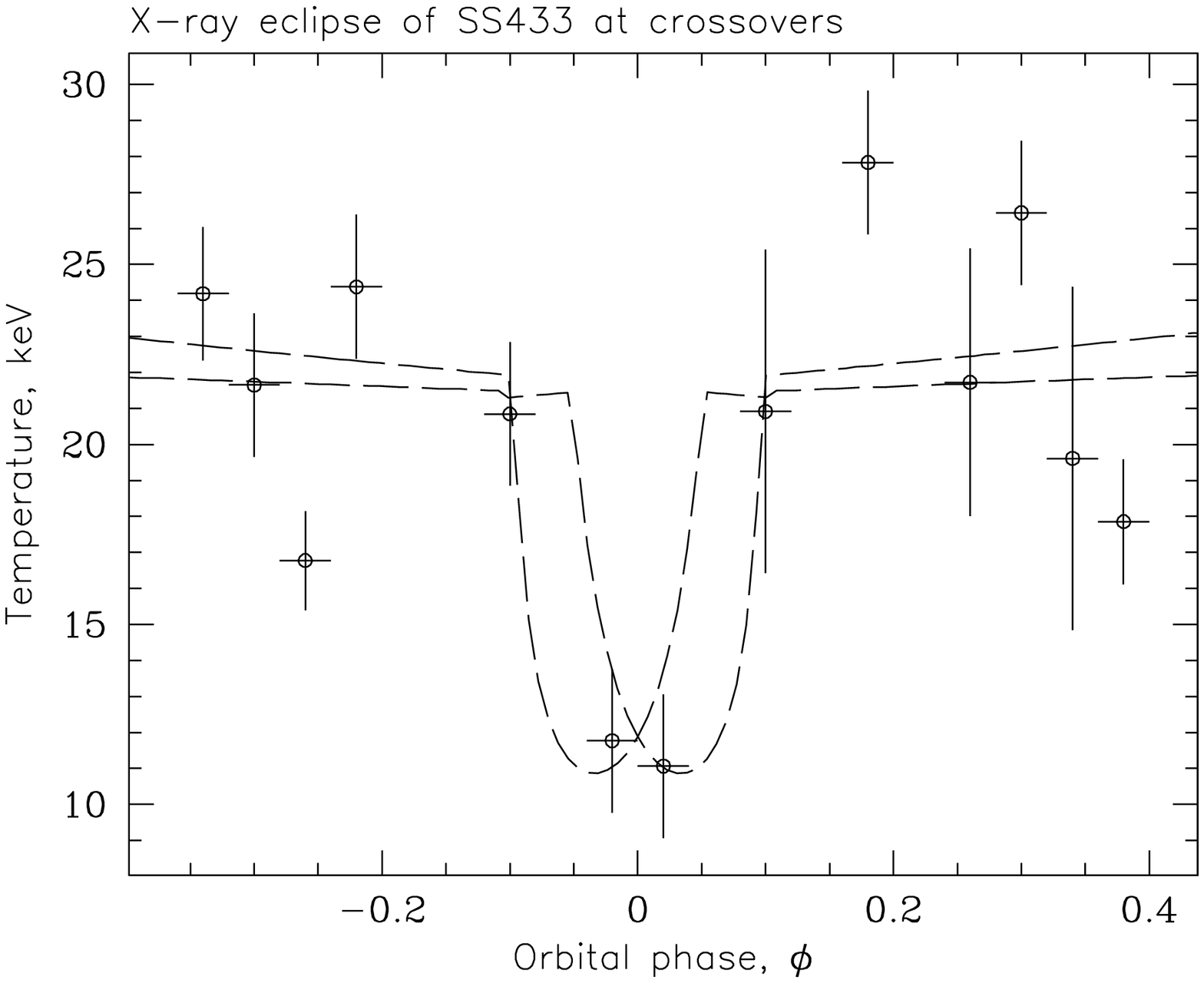}{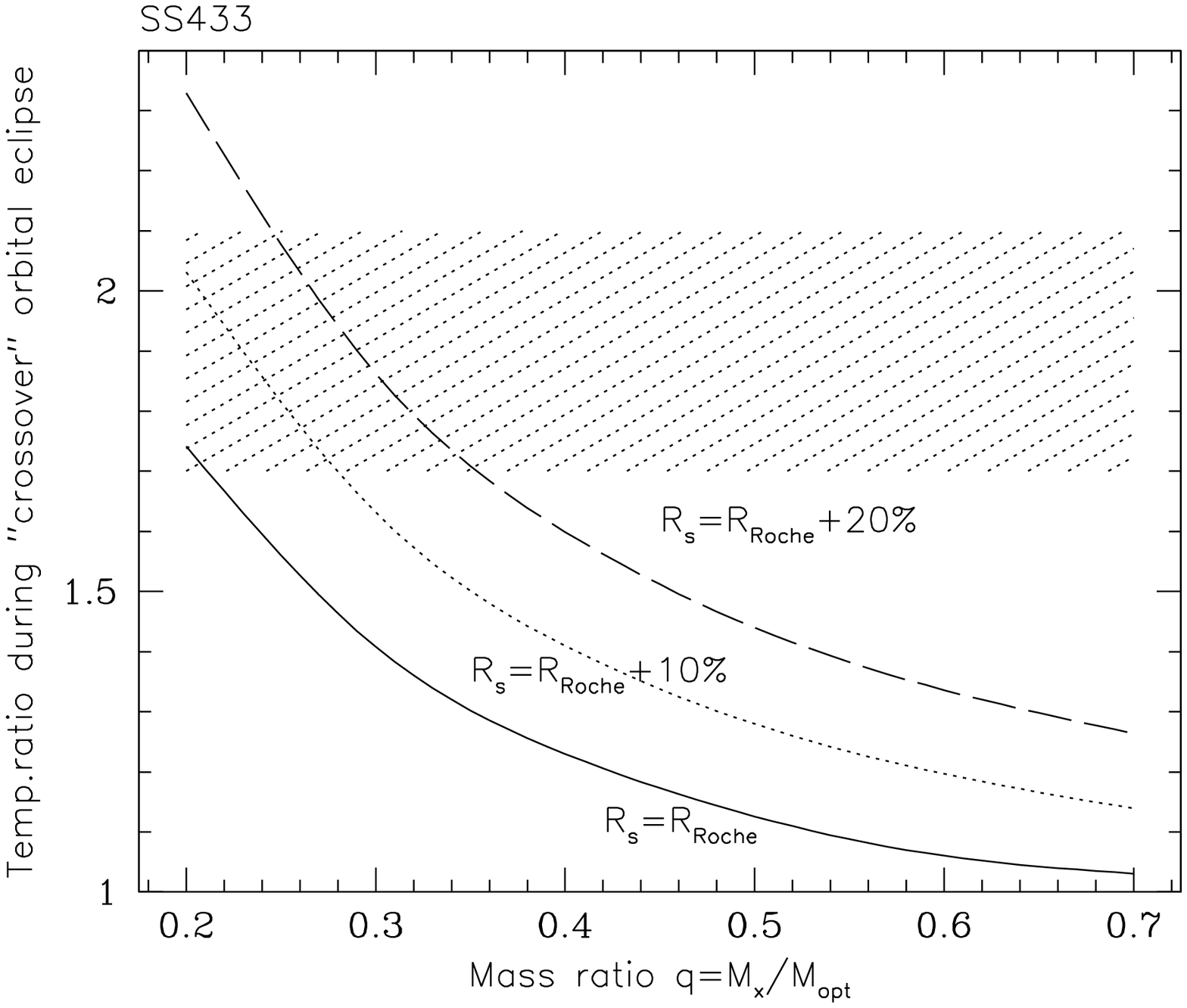}
\caption{Left panel: orbital X-ray eclipse observed at the
precessional phases $\psi\sim0.33,0.66$. Right panel: the ratio of the maximum visible jet temperatures during the
orbital eclipse at these precessional phases
 as a function of the mass ratio $q$.  The hatched area shows the observational
constrains.}
\end{figure}

In Fig.4(left panel) we present the profile of the maximum
X-ray temperature measured in the orbital eclipse near
precessional phases $\psi\sim0.33,0.66$. The
orbital X-ray eclipse is clearly visible and shows an appreciable
depth. The ratio of the maximum jet temperature derived from
off-eclipse observations to that of the eclipsed jet is
$\sim1.9\pm0.2$.  Dashed lines in the figure show examples of eclipsing profiles
obtained in our model. They are not symmetric due to the different jet inclination with respect to the binary
 orbital plane at this phases.

The ratio of the maximum jet temperatures in the eclipse and off the eclipse
is plotted in Fig.4 (right panel) as a function of the
binary mass ratio $q$. The solid curve shows the ratio obtained
under the assumption that the size of the star equals to the volume averaged
radius of the Roche lobe.

We have shown in Section 2 that the size of the eclipsing region might
be large then the Roche lobe of the optical star.
The innermost parts of dense stellar wind also can absorb X-rays thus
increasing the duration and depth of the X-ray eclipse. So in Fig. 4 (right panel) we also
plot by the dotted and dashed curves
the model ratio of maximum jet temperatures in- and out- of the
eclipse assuming the star radius to be $R=1.1 R_{\rm Roche}$ and
$R=1.2R_{\rm Roche}$ correspondingly.

 From Fig.4 (right panel) we can conclude that the
binary mass ratio in SS433 should not be significantly larger
than $q\sim0.3-0.35$, because in that case the optical star would be
too small to provide the observed depth of the eclipse or the radius of the star
(i.e. the radius of the X-ray eclipser  which can exceed
the actual star radius) should be larger than $R/a\ga 0.5-0.55$.

\section{Conclusion}
 Using extensive observational
data obtained by RXTE, we studied systematic variations of X-ray
emission of SS433 caused by precessional and orbital motions in the
system.

We compared X-ray spectra of SS433 near and in the eclipse and
obtained strong signatures of a significant ($N_{H}>10^{23}$
cm$^{-2}$) photoabsorption of X-rays near the companion star. We
argue that this may be due to the presence of a dense stellar wind
from the companion star.   The size of the X-ray eclipsing region can be
significantly larger than the  size of the Roche lobe due to absorption in
the inner stellar wind.

We recovered the temperature profile of plasma along the jet and estimated the jet cross
section at its basement. The maximum visible temperature is found to be
$T\sim30$ keV at a distance of $l/a=0.06-0.09$ from the compact object.
The radius of the jet at
this distance is $r_0/a=0.01-0.016$.

We reliably detected orbital eclipses in the RXTE/PCA observations
of SS433 during the ``crossover'' precessional phases when the X-ray
jets and the axis of the accretion disk lie exactly in the plane of
the sky. The observed depth of the X-ray eclipse implies that the
size of the star is larger than $R/a\ga0.5$, yielding an upper bound
on the mass ratio of the components in SS433 $q<0.3-0.35$ assuming
that the radius of the eclipser (star plus inner wind) can not be
much larger than $1.2R_{\rm Roche, secondary}$ .

\section{Acknowledgements}

Research has made use of data obtained from High Energy Astrophysics
Science Archive Research Center Online Service, provided by the
NASA/Goddard Space Flight Center. This work has been supported by
RFBR grants N\,04-02-16349, N\,05-02-19710, N\, 06-02-16025, N\, 04-02-17276 and N\,
04-02-16720, and by joint RFBR/JSPS grant N\,05-02-12710.

\begin {references}
\reference Achmad, L., Lemers H.J.G.L.M., \& Pasquini, ~L. 1997, A\&A, 320, 196

\reference Fabrika, S. 2004, Astrophysics and Space Physics Reviews, 12, 1

\reference Koval', E. \& Shakura N., 1989, in Proc. 23rd ESLAB Symp. on Two-Topics in X-ray Astronomy (ESA SP-296), 479

\reference Marshall, H., Canizares, C., \& Schulz, N. 2002, \apj, 564, 941

\reference Paczynski B., 1977, \apj, 216, 822

\end {references}
\end{document}